# Enhanced Terahertz Spectroscopy of a Monolayer Transition Metal Dichalcogenide

*Xin Jin[#], Vincenzo Aglieri[#], Young-Gyun Jeong, Atiye Pezeshki, Lilian Skokan, Mostafa Shagar, Yuechen Jia, Pablo Bianucci, Andreas Ruediger, Emanuele Orgiu, Andrea Toma, and Luca Razzari\**

X. Jin, Y.-G. Jeong, A. Pezeshki, L. Skokan, M. Shagar, Y. Jia, A. Ruediger, E. Orgiu, L. Razzari
Institut National de la Recherche Scientifique, Centre Énergie Matériaux Télécommunications (INRS-EMT), 1650 Boulevard Lionel-Boulet, Varennes, Quebec, J3X 1P7, Canada
E-mail: luca.razzari@inrs.ca

V. Aglieri, A. Toma
Clean Room, Istituto Italiano di Tecnologia, via Morego 30, Genova, 16163, Italy

P. Bianucci
Department of Physics, Concordia University, 7141 Sherbrooke Street West, Montreal, Quebec, H4B 1R6, Canada

[#]These authors contributed equally to this work.

**Abstract:** Two-dimensional materials, including transition metal dichalcogenides, are attractive for a variety of applications in electronics as well as photonics and have recently been envisioned as an appealing platform for phonon polaritonics. However, their direct characterization in the terahertz spectral region, of interest for retrieving, *e.g.*, their phonon response, represents a major challenge, due to the limited sensitivity of typical terahertz spectroscopic tools and the weak interaction of such long-wavelength radiation with sub-nanometer systems. In this work, by exploiting an *ad-hoc* engineered metallic surface enabling a ten-thousand-fold local absorption boost, we perform enhanced terahertz spectroscopy of a monolayer transition metal dichalcogenide (tungsten diselenide) and extract its dipole-active phonon resonance features. In addition, we use these data to obtain the monolayer effective permittivity around its phonon resonance. Via the direct terahertz characterization of the phonon response of such two-dimensional systems, this method opens the path to the rational design of phonon polariton devices exploiting monolayer transition metal dichalcogenides.

**1. Introduction**

Phonon polaritonics, dealing with the hybridization of lattice vibrations with electromagnetic radiation, has become a rapidly growing field that provides versatile solutions for applications requiring manipulation and control of long wavelength light (encompassing the mid-infrared (IR) and terahertz (THz) ranges).[1] This is typically realized in bulk materials that support the emergence of surface phonon polaritons (PhPs) in the spectral region – known as the Reststrahlen band – between their transversal optical (TO) and longitudinal optical (LO) phonon modes. Compared to plasmon polaritons, PhPs feature lower intrinsic loss and better light confinement for long wavelengths.[2] They have thus been proposed for a variety of applications in IR and THz photonics, *e.g.*, in surface-enhanced IR absorption spectroscopy,[3] heat management,[4] radiative cooling,[5] imaging and sensing.[6]

In this context, two-dimensional (2D) materials are becoming game changers that enable PhP responses from atomically thin layers.[2b, 7] For example, it has been shown that PhPs in 2D multilayered van der Waals materials such as hexagonal boron nitride (hBN)[8] and α-phase molybdenum trioxide ($\alpha$-$MoO_3$)[9] can exhibit a natural hyperbolic dispersion (related to the anisotropy in the sign of the principal permittivity components), which in turn makes it possible to achieve unique functionalities like nanoscale directional energy transfer and subdiffractional focusing. Single layers of 2D materials can further push PhP confinement to its ultimate limits, as it has been observed in monolayer hBN,[10] where the PhP wavelength can be hundreds of times smaller than that of light of the same frequency in free space.

An interesting class of 2D material systems is represented by transition metal dichalcogenides (TMDs). Thanks to their semiconducting properties, TMDs have the potential to be employed in next-generation high-performance, low-power-consumption electronic and computing technologies.[11] Furthermore, TMDs exhibit a direct bandgap in their monolayer form and host tightly bound excitons that are stable even at room temperature, which makes them appealing for optoelectronic applications.[12] TMDs also possess the attractive advantage of gate-tunable optical properties, which is not available in insulating platforms like hBN.

For what concerns long-wavelength photonics, semiconducting TMDs hold promise in relation to their distinctive characteristics, for example regarding hyperbolic PhPs,[1, 2b, 13] chiral phonons,[14] and potential for ultraslow polariton interactions.[2b] Indeed, although slow-moving

PhPs in TMDs are not ideal for propagation-related (*e.g.*, waveguiding) applications, such low-velocity modes are expected to have long lifetimes, guaranteeing an extensively prolonged energy exchange with, e.g., interacting molecules, which can benefit strong light-matter coupling experiments.[2b]

As a member of the semiconducting TMD family, tungsten diselenide ($WSe_2$) is considered a promising platform for electronics as well as optoelectronics,[15] and has recently shown long-range ferromagnetic order at room temperature when doped with vanadium,[16] thus providing interesting opportunities also for spintronics. Optical phonons play an important role in all these applications, as they are involved in strong electron-phonon scattering processes in TMDs,[17] which can ultimately limit device performance.

Despite such a significant interest, dipole-active phonon modes of monolayer TMDs are so far explored only via first-principles calculations or indirectly, *e.g.*, via Raman characterization, due to evident difficulties in performing direct spectroscopic measurements on such ultra-thin material systems in their PhP regions (*i.e.*, in the THz range). This also implies that a reliable estimation of their dielectric permittivities at such frequencies, essential for the design of PhP devices, is still unavailable.[1] In this work, by exploiting a nano-engineered metallic surface featuring properly-shaped nanoslot resonators as a platform for enhanced THz spectroscopy, we successfully characterized the THz response of monolayer $WSe_2$ in the spectral region of its E' optical phonon mode. By fitting the experimental data with the results of electromagnetic simulations, we were able to estimate the E' phonon resonance frequency, lifetime, and group velocity, as well as extract the effective permittivity of monolayer $WSe_2$ in the THz range. Our method can be directly extended to other monolayer TMDs and, more broadly, to other 2D material platforms, for the THz characterization of their optical phonon response and the advanced design of 2D THz PhP devices.

## 2. Results and Discussion

### 2.1. Design, Fabrication, and THz Characterization of X-shaped Nanoslot Array

To perform enhanced THz spectroscopy of the targeted TMD monolayer, we designed a substrate featuring resonant metallic nanostructures to boost the local THz electric field, in analogy with the strategy we previously exploited to characterize[18] (or manipulate[19]) the phonon response of semiconducting nanocrystals. While in these former studies rod-like nanoantennas were employed, our new design makes use of a resonant nanoslot geometry (nanoslots are apertures with nanoscale features on a metallic surface[20]), to provide better mechanical support to the 2D material. Resonant nanoslots typically present an area at their center where the local electric field is enhanced (see Figure S1 in Supporting Information). However, the enhancement factor in this area can be relatively weak. To improve such enhancement and, at the same time, preserve a robust mechanical support that avoids excessive stretching and bending of the 2D material, we adopted a heuristic approach that exploited the tapering of the nanoslots while limiting the size of the apertures.

Recently, we have shown that tapering of rod-like THz nanoantennas reduces the Ohmic loss of such resonators, thus improving their electric field localization properties.[21] This technique is also effective for "negative" structures such as the nanoslots considered here. In this case, a straightforward geometry to better confine the electric field at the nanoslot center is represented by a bowtie-shaped slot (i.e., a structure tapered at the two extremities) with an extended nanogap area. Our final design eventually takes on a characteristic X shape (**Figure 1**) that allows us to further enhance the electric field in the nanogap and, at the same time, limit the size of the individual apertures composing the structure (for a comparison among a uniform rectangular nanoslot, two examples of bowtie nanoslots, and the X-shaped nanoslot, see Figure S1 in Supporting Information).

Figure 1a shows a sketch of the investigated array of X-shaped nanoslots prepared in a 100-nm-thick gold layer on a silicon substrate, highlighting its key design characteristics. For the array to have a local field resonance in the proximity of the targeted E' phonon mode of monolayer $WSe_2$, which is located at around 7.5 THz (as estimated by previous first-principles studies and Raman measurements),[22] the X-shaped structure features a gap in its center of $w$ (500 nm) × $g$ (50 nm) along the $x$ and $y$ directions, respectively, while the four arms of the X shape have a height $h$ = 2.2 μm along the $y$ direction (see additional geometrical details of the X shape in the bottom right panel of Figure 1a). The array periodicity in the $x$ and $y$ directions is set to $s_x = s_y$ = 5.4 μm in a square lattice geometry.

The designed array was fabricated using electron-beam lithography and covers an area of 2.5 × 2.5 mm² (for fabrication details, see Methods). Figures 1b,c show scanning-electron-microscope (SEM) images of the fabricated X-shaped nanoslot array (for a close-up zoom of the nanogap area, see also Figure S2 in Supporting Information). To characterize the THz response of the fabricated structure, we used a Fourier transform spectrometer in a transmission configuration (see THz characterization details in Methods). The measured transmission spectrum of the bare array is shown in Figure 1e (grey curve), highlighting a far-field resonance position at around 7.3 THz.

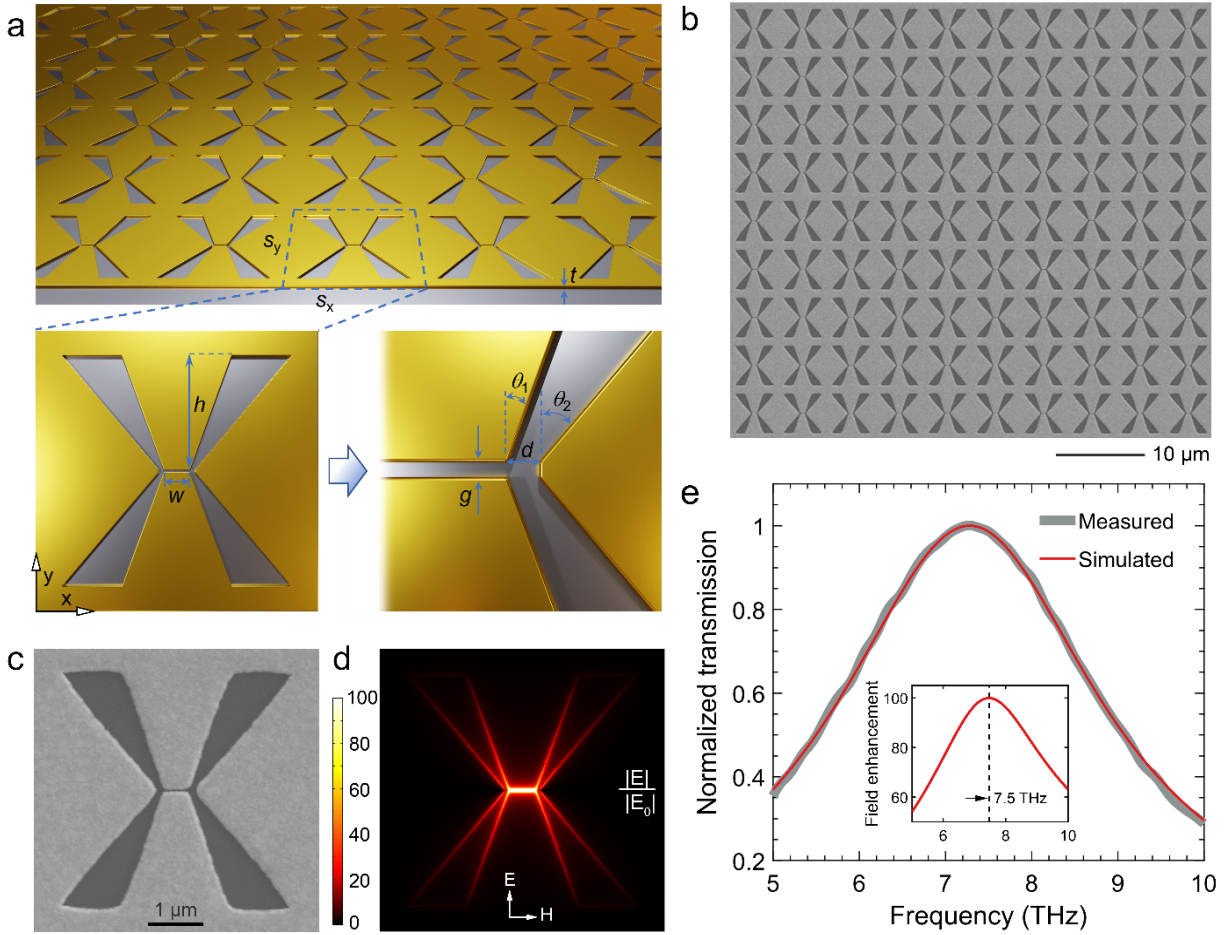

**Figure 1.** X-shaped nanoslot array: design, fabrication, and THz characterization. a) Schematic of the X-shaped nanoslot array on a silicon substrate. Relevant parameters: $s_x = s_y = 5.4$ μm; $t = 100$ nm; $h = 2.2$ μm; $w = 500$ nm; $g = 50$ nm; $d = 100$ nm; $\theta_1 = 20°$; $\theta_2 = 40°$. b) Scanning electron microscope (SEM) image of the fabricated array. c) A zoomed-in SEM image (5.4 × 5.4 μm$^2$) of a single nanoslot. d) Spatial distribution of the electric field enhancement on a plane 0.65 nm above the metallic surface, at the frequency of 7.5 THz. $E_0$ is the electric field amplitude of the incident THz light, while E is the local electric field amplitude. e) Measured (grey curve) and simulated (red curve) transmission spectra of the X-shaped nanoslots array (the incident THz light is polarized along the *y*-axis). Inset: Electric field enhancement of the X-shaped nanoslots as a function of frequency, numerically estimated at the nanogap center on a plane 0.65 nm above the metallic surface (which corresponds to the thickness value of monolayer WSe$_2$[34]).

All numerical simulations for the sample design were performed using COMSOL Multiphysics. In addition, after the THz characterization of the fabricated device, we employed the Nelder-Mead optimization method incorporated in the COMSOL numerical solver to best fit the measured array transmission (see details in Methods). To this purpose, the permittivity of gold in the THz range was described via the Drude equation: $\varepsilon_D(\nu) = 1 - \nu_p^2 / (\nu^2 + i\gamma_D\nu)$, where $\nu$ is the frequency, $\nu_p$ = 2080 THz the plasma frequency of gold taken from literature,[23] while $\gamma_D$ is the damping factor, which we used as the key fitting parameter in the simulations. The result of the fitting procedure is shown in Figure 1e (red curve), returning a value for the damping factor $\gamma_D$ = 449 THz. The simple adjustment of the damping value allows us to obtain an excellent agreement between the experimental transmission spectrum of the nanoslot array in the THz range and the simulated one, which will be instrumental in properly extracting the permittivity of monolayer $WSe_2$ (see below).

Numerical simulations also enable the visualization of the electric field right above the X-shaped nanoslot, where the 2D material will be located. Such local field is highly concentrated in the nanogap region (see Figure 1d), and its resonance is well aligned to the targeted dipole-active phonon mode of monolayer $WSe_2$ at 7.5 THz, with a peak enhancement value estimated to be $|E|/|E_0|$ ~100 at the gap center, as shown in the inset of Figure 1e. Further details regarding the electric field distribution along the $z$-axis, tunability of the resonance, and polarization dependence of the X-shaped nanoslot array are reported in Supporting Information (Figure S3-S5).

## 2.2. Phonon Response of Monolayer $WSe_2$: First-Principles Calculations, Raman Spectroscopy, and Direct THz Characterization

Phonons in monolayer $WSe_2$ can be investigated through first-principles calculations.[22, 24] Using this approach, we computed its phonon dispersion by means of the calculation package Quantum Espresso (see configurations in the Methods section),[25] which incorporates, among others, methods such as density functional theory (DFT)[26] and density functional perturbation theory (DFPT).[27] **Figure 2**a shows the result of this calculation, while Figure 2b illustrates the atomic displacement of two characteristic phonon modes in the monolayer (the out-of-plane Raman-active $A_1'$ mode and the in-plane E' mode, which is both dipole- and Raman-active) and their connection with the phonon modes of bulk $WSe_2$. Examining the phonon dispersion in Figure 2a, we can first notice the absence of LO-TO splitting at the Γ point for the E' mode (located at ~7.5 THz, i.e., ~250 cm$^{-1}$), a distinctive feature of 2D polar materials (such as the TMDs) in their monolayer form.[28] We can also see that the $A_1'$ mode closely approaches the $E'$ mode at the Γ point. The close spectral proximity between these two modes (*i.e.*, $E'$ and $A_1'$) is characteristic of monolayer $WSe_2$, while in multilayer systems the respective modes tend to progressively separate in frequency when increasing the number of layers.[29]

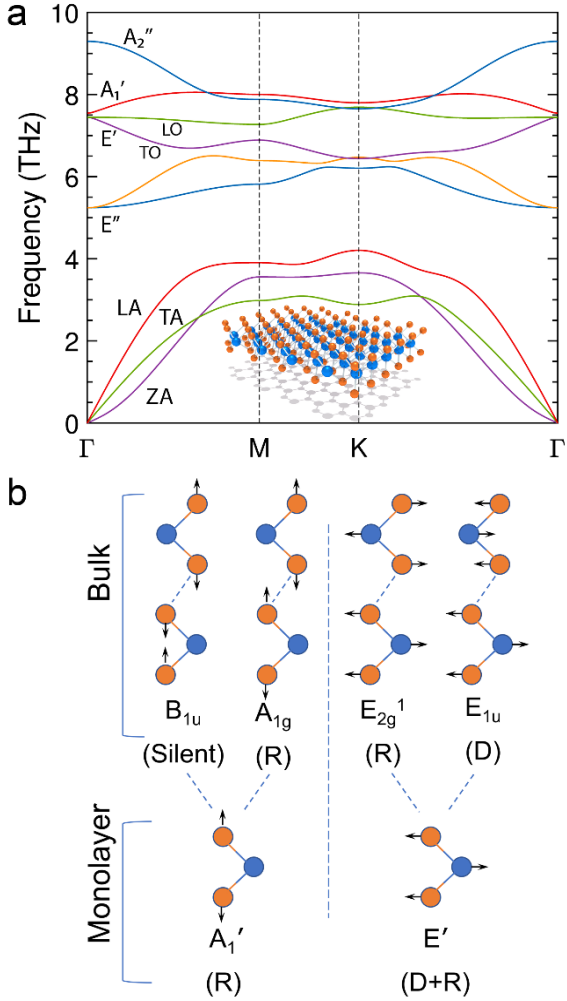

**Figure 2.** Phonon modes in monolayer $WSe_2$. a) Calculated phonon dispersion of monolayer $WSe_2$. b) Schematic of atomic displacements for the optical phonon modes $A_1'$ and $E'$ of monolayer $WSe_2$ and their bulk mode counterparts. Notation: Raman-active (R), dipole-active (D), both Raman and dipole inactive (Silent).

In our experiments, we employed a monolayer $WSe_2$ (*2Dsemiconductors Inc.*) prepared via chemical vapor deposition (CVD) and then transferred on the X-shaped nanoslot array. **Figure 3**a shows an optical image of the sample, where it is possible to identify the array at the center of the gold-coated silicon chip, as well as the $WSe_2$ layer (darker area) that fully covers the nanoslots (see also the zoom-in picture in the right panel of Figure 3a). Monolayer $WSe_2$ exhibits a direct bandgap with a peak photoluminescence (PL) emission at ~1.65 eV, while from monolayer to multilayer the bandgap becomes indirect and narrows (with PL peak at 1.54 eV for the bilayer, 1.46 eV for the trilayer).[30] PL can thus be used to identify the number of WSe$_2$ layers. In this regard, we performed PL emission spectroscopy on our sample at different locations over the array, by employing a 445-nm laser excitation. As the seven measurements reported in Figure 3b show, the PL spectra of the CVD-grown WSe$_2$ transferred over the X-shaped nanoslot array are consistently peaked at around 1.65 eV, confirming the monolayer nature of the employed TMD over the area of interest.

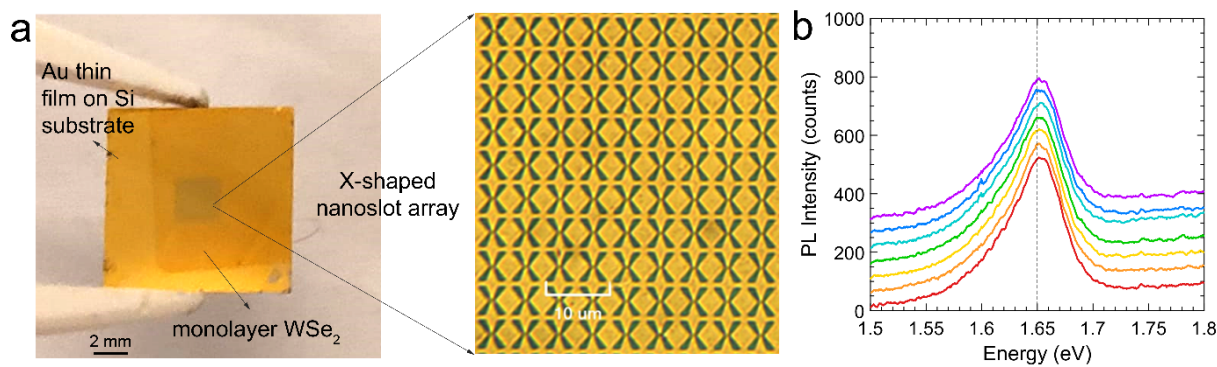

**Figure 3.** Sample image and photoluminescence measurements. a) Optical image of the X-shaped nanoslot sample covered with monolayer WSe$_2$ (darker area in the left panel). b) Photoluminescence spectra of the WSe$_2$ layer, taken at seven different points over the surface of the X-shaped nanoslot array. Curves are shifted vertically for clarity.

To experimentally identify the spectral location of the E' mode of the monolayer WSe$_2$ covering our sensing chip, we performed micro-Raman spectroscopy measurements, exploiting the fact that this mode is also Raman active. Since, as mentioned earlier, the E' and A$_1$' modes almost overlap in monolayer WSe$_2$, we employed polarized Raman spectroscopy, given that the A$_1$' mode dominates the Raman response in a parallel-polarization configuration (i.e., a configuration in which the scattered light is collected via a polarizer with orientation parallel to the input laser polarization) but is almost forbidden in crossed polarization, while the E' mode contributes nearly equivalently in both cases.[29b, 31] This can be quantitatively seen via the calculation of the Raman activity in the two configurations (**Figure 4**a). 99.7% and 57.1% of Raman activity of the A$_1$' and E' modes are found in the parallel-polarization configuration (blue curve), respectively. As a result, the E' mode, with its 42.9% residual activity, is dominant in the cross-polarization configuration (red curve) with a negligible A$_1$' contribution (see also the isolated contributions and amplitudes of the two modes for the two polarization configurations in Figure S6 of Supporting Information). Figure 4b presents the example of a polarized Raman spectroscopy measurement taken at a nanoslot location (laser excitation: 473 nm, power: 5.91 mW, focal spot size: 0.6 μm, acquisition time: 700 s), which results to be in good agreement with the calculation outcomes (Figure 4a) for what concerns the WSe$_2$ Raman peak locations (for additional unpolarized Raman measurements and related comments, see Figure S7 of Supporting Information). This investigation thus confirms the spectral location of the E' phonon mode for our sample (~250 cm$^{-1}$, i.e., ~7.5 THz), which is the target of our THz characterization study.

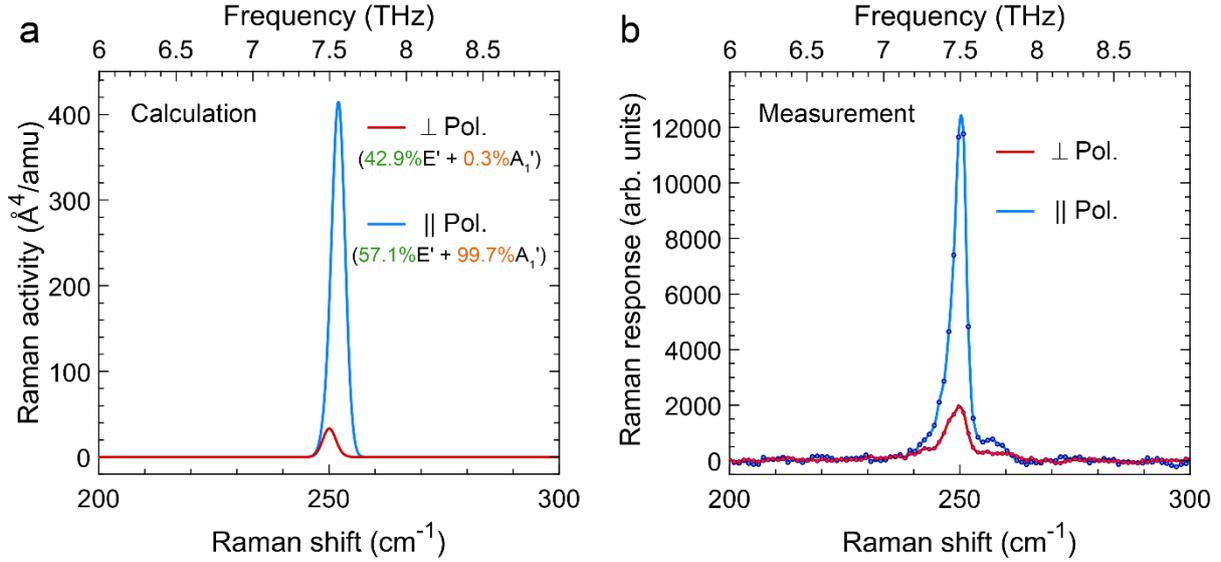

**Figure 4.** Polarized Raman response of monolayer WSe$_2$. a) First-principles calculation of the Raman activity of monolayer WSe$_2$ for a parallel (blue curve) or crossed (red curve) polarization configuration. A Gaussian broadening of 2 cm$^{-1}$ is used for visualization purposes. b) Experimental polarized Raman spectra of the WSe$_2$ layer, taken at an X-shaped nanoslot gap area by employing a parallel (blue curve) or crossed (red curve) polarization configuration.

**Figure 5** (grey curve) shows the THz transmission measurement performed on the X-shaped nanoslot array covered with the monolayer WSe$_2$. As can be seen, an anti-resonant dip forms on the array response at around 7.5 THz (i.e., in the spectral position of the WSe$_2$ E' mode). This spectral behavior is typical of nanoantenna-enhanced vibrational spectroscopy measurements[18, 32] and results from the coupling between the resonant mode of the employed sensing nanostructures and the targeted vibrational mode. Our technique is thus capable of retrieving the dipole-active phonon spectral signature of monolayer WSe$_2$, by means of the field enhancement provided by the nanoslots. To the best of our knowledge, this is the first experimental report of direct far-field THz phonon spectroscopy of an atomically thin material. Making use of this measurement, we can finally extract an effective permittivity for the investigated monolayer WSe$_2$ and also evaluate its key phonon resonance features, as detailed in the following section.

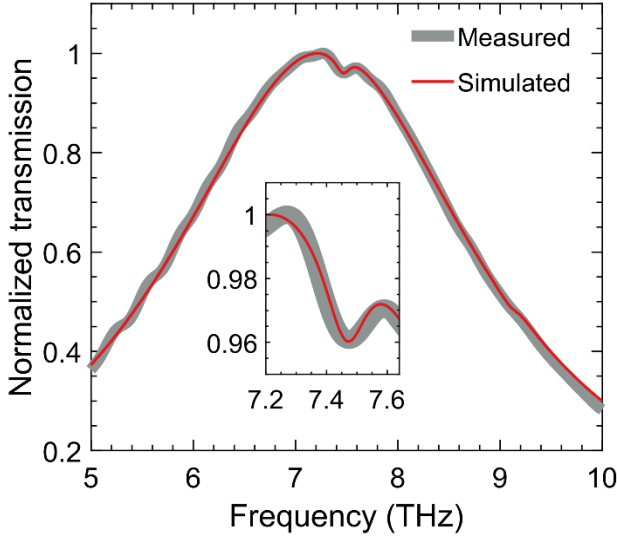

**Figure 5.** Enhanced THz spectroscopy of monolayer WSe$_2$. Measured transmission spectrum of the X-shaped nanoslot array covered with monolayer WSe$_2$ (grey curve). The red curve is the simulated transmission spectrum obtained via the numerical fitting of the experimental data, by varying the permittivity of monolayer WSe$_2$. Inset: Zoomed-in plot showing the actual fitting range.

### 2.3. Permittivity Extraction

As recent literature has shown, even in the absence of LO-TO splitting, monolayers of polar materials support a strongly confined PhP mode that corresponds to the 2D LO phonon.[10b, 28, 33] In particular, ref.[28] introduced the convenient idea of utilizing, for compatibility with standard numerical tools, a "bulkified" medium (with thickness $l$ equal to the monolayer thickness) to reproduce the monolayer response in electromagnetic simulations. The effective in-plane permittivity of such medium (its out-of-plane permittivity is taken as equal to 1) can be written as:

$$\varepsilon_{\text{eff}} = 1 + \frac{1}{l}\frac{4\varepsilon_{\text{env}}v_0 u_{\text{g}}}{v_0^2 - v^2 - i v\tau^{-1}}, \qquad (1)$$

where $\varepsilon_{\text{env}}$ is the permittivity of the surrounding medium and $u_g$, $v_0$, and $\tau$ are the LO phonon group velocity, resonance frequency, and lifetime, respectively. Using this model, we were able to employ numerical simulations to best fit the experimental transmission spectrum of Figure 5 (red curve, see details in Methods), in a manner analogous to what was previously done for the bare array transmission. In the simulations, the layer thickness was set to $l$ = 0.65 nm, considering the reported interlayer spacing value for WSe$_2$,[34] while $\varepsilon_{env}$ = 1, taking into account that the monolayer is effectively suspended in air in the hot spot regions of the nanoslot array (*i.e.*, the nanogap areas), where spectroscopy is enhanced (see Figure S8 in Supporting Information). Through the fitting procedure, we independently retrieved the values of the LO phonon frequency, lifetime, and group velocity, which are respectively associated with the spectral position, linewidth, and relative transmission change of the anti-resonant feature observed in the experimental data (Figure 5). We thus obtained $v_0$ = 7.47 THz, in good agreement with the results of both first-principles calculations and Raman measurements, a long phonon lifetime $\tau$ = 5.56 ps, and an ultraslow group velocity $u_g$ = 610 m/s. Regarding the latter quantity, such experimentally estimated value of the phonon group velocity at the Γ point gives valuable insight into the small-wavevector dispersion of 2D LO phonons in WSe$_2$ and is found to be fairly aligned with previously reported theoretical calculations (1600 m/s[35]).

**Figure 6** a,b show the dispersion of the real and imaginary part of the effective permittivity and refractive index of monolayer WSe$_2$, as extracted from our experiments. Making use of these data in electromagnetic simulations, we can estimate the visibility enhancement of the phonon spectral signature of monolayer WSe$_2$ when placed on the X-shaped nanoslot array. Figure 6c shows the transmission spectra with (darker color curves) and without (lighter color curves) monolayer WSe$_2$, in the case in which silicon is directly used as a substrate (simulated blue curves), or when the nanoslot array is employed (red curves: simulation; grey curves: experiment). For comparison purposes, all curves are normalized to the transmission of a bare silicon substrate (see further details on how these curves are obtained in Methods). As can be seen, the overall experimental transmission of the nanoslot array is approximately 50% of the simulated one. This reduction in transmission is due to the fact that, owing to fabrication constraints, the patterned nanoslot area in the gold film (see Figure 1a) is smaller than the illuminating THz beam in our experiment, whereas in the simulations an infinitely extended array is considered through periodic boundary conditions (the origin of the reduced transmission was confirmed via the measurement of a gold-coated silicon sample with a square aperture of the same size as the X-shaped nanoslot array, *i.e.*, 2.5 × 2.5 mm$^2$ – see Figure S9 in Supporting Information). By calculating the transmission difference $\Delta T$ with and without monolayer WSe$_2$, we can directly compare the visibility of the E' phonon resonance in the various cases presented in Figure 6c. Figure 6d shows that $\Delta T$ is extremely small (less than 0.1%) when monolayer WSe$_2$ is placed directly on a silicon substrate (blue curve). Conversely, it increases by about 10 times in the simulations when using the X-shaped nanoslot array (red curve). Solely due to the limited size of the fabricated array, the experimental visibility enhancement is instead ~5 (see grey curve), a value that nonetheless allows the phonon resonance peak to emerge from the noise floor of our measurement. The origin of the visibility enhancement is the increased THz absorption of WSe$_2$ in the nanogaps of the X-shaped nanoslots. Figure 6e shows the simulated THz absorption density evaluated in the center of the nanoslot gap at half the thickness of the WSe$_2$ layer (red curve), as well as the same quantity for a WSe$_2$ layer directly placed on silicon (blue curve). A significant $10^4$ increase in the local absorption of WSe$_2$ is found in the nanogaps, which is key to enabling the direct observation of the phonon spectral signature of such an atomically thin material in the THz range.

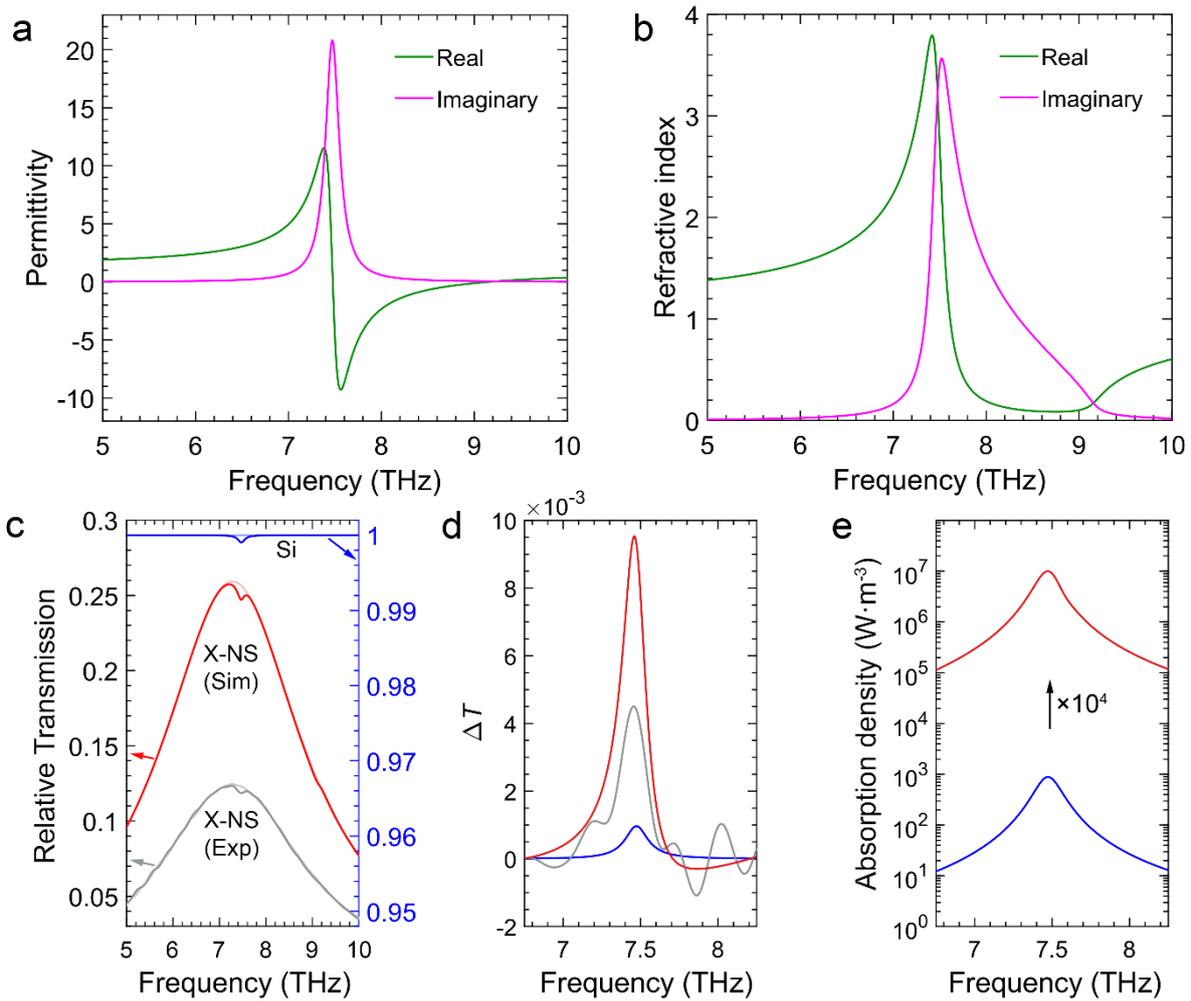

**Figure 6.** Effective permittivity of monolayer WSe$_2$, phonon resonance visibility and local absorption enhancement. a) Effective permittivity of monolayer WSe$_2$, as extracted via the procedure described in the text (real part: green curve; imaginary part: magenta curve). b) Corresponding refractive index. c) Relative transmission with (darker color curves) and without (lighter color curves) monolayer WSe$_2$ for the case of a silicon substrate (simulated blue curves) and the X-shaped nanoslot array (X-NS – red curves: simulation; grey curves: experiment);

simulated curves are estimated using the permittivity shown in (a). The scale of the right axis (referring to the silicon substrate case) is expanded by a factor of 5 compared to that of the left axis, for comparison purposes. d) Transmission difference with and without monolayer WSe$_2$ (a quantity used to estimate the phonon peak visibility), for the three cases reported in (c). e) Simulated THz absorption density evaluated in the middle of the WSe$_2$ layer, when the latter is positioned on silicon (blue curve) or over the nanogap of an X-shaped nanoslot (red curve).

## 3. Conclusion

We have performed direct THz spectroscopy of monolayer WSe$_2$ around its E' optical phonon mode, by exploiting the local field enhancement of an *ad-hoc* designed array of X-shaped nanoslots. Such resonant nanostructures have enabled a $10^4$ boost in the local THz absorption of the TMD material, resulting in a 5-fold increase (10-fold theoretical) in the visibility of its phonon spectral signature. This has allowed us to retrieve the effective permittivity of monolayer WSe$_2$ in the range 5-10 THz, as well as its key optical phonon characteristics, i.e., the phonon resonance frequency $\omega_0 = 7.47$ THz, lifetime $\tau = 5.56$ ps, and group velocity at the $\Gamma$ point $v_g = 610$ m/s. Such long phonon lifetime and slow group velocity, the latter directly obtained from experiments for the first time (to the best of our knowledge), highlight the possibility of exploiting PhPs in this 2D material platform for, e.g., nanolocalized sensing, strong radiation-matter interactions, or the exploration of forbidden transitions in the THz range.[36]

Our strategy can be straightforwardly applied to other monolayer TMDs as well as different 2D materials. It can also enable the investigation of few-layer heterostructures, to explore the tailoring of the THz optical phonon response in these artificially created stacks. Furthermore, inverse design[37] can be used to define the sensing substrate geometry for an optimized spectroscopic enhancement. While in this study, a CVD-grown monolayer was employed, and our method can also be extended to the characterization of mechanically exfoliated samples, making use of recently developed techniques for (i) the exfoliation of large-area flakes (mm-size and beyond),[38] which are directly compatible with measurements on THz resonator arrays, or (ii) the selective excitation of individual metallic resonators in the THz range,[39] which could be exploited to sense flakes of more typical (micrometric) size. Such an extensive range of opportunities for the extraction of the THz optical properties of 2D materials thus promises to open up unprecedented pathways for the design of ultrathin PhP devices operating in the THz spectral region.

## 4. Methods

*Electromagnetic simulations and fittings:* Electromagnetic simulations were performed using COMSOL Multiphysics,[40] exploiting the Optics Module and the Optimization Module. In simulations, a lattice of X-shaped gold nanoslots was placed on a silicon substrate. The top air and bottom silicon domains were truncated by two perfectly matched layers to avoid any spurious reflections. Along the *x*- and *y*-directions, two pairs of Floquet periodic boundary conditions were used to simulate an infinitely large array. The refractive index of silicon was set to 3.42. The permittivities of gold and WSe$_2$ in the THz range were described via the models detailed in the main text. Regarding the illumination condition, light with *y*-axis polarized electric field from the air domain was employed to excite the nanoslots. Using COMSOL finite-element-method solver in the frequency domain, near- and far-field quantities could be obtained, such as the local electric field enhancement, absorption, and scattering cross-sections, as well as the integral average of the power flow for calculating the array reflection and transmission.

For the numerical fittings, the Nelder-Mead optimization method incorporated in the numerical solver was used to fit the measured transmission curves. In the case of the bare X-shaped nanoslot array, the objective function was defined as $f_{\text{obj}} = \sum_{\nu=5\text{ THz}}^{10\text{ THz}} |T_{\text{sim}}(\nu, \gamma_D) - a_0 \cdot T_{\text{exp}}(\nu)|$, in which $a_0$ is a normalization parameter, $\gamma_D$ is the damping factor in the Drude equation describing the permittivity of gold, and $T_{\text{sim/exp}}$ is the simulated or measured relative transmission spectrum (calculated as the transmission ratio of the nanoslot array sample to a reference silicon substrate). $a_0$ and $\gamma_D$ were set as the fitting parameters to search for the minimum of $f_{\text{obj}}$ in the range 5 – 10 THz. The fitting procedure, performed exploiting the Béluga cluster of the Digital Research Alliance of Canada, returned the values $a_0 = 2.1486$ and $\gamma_D = 449$ THz. In the case of the array covered with monolayer WSe$_2$, the objective function was defined as $f'_{\text{obj}} = \sum_{\nu=7.2\text{ THz}}^{7.64\text{ THz}} |T'_{\text{sim}}(\nu, \nu_0, u_g, \tau) - a_1 \cdot T'_{\text{exp}}(\nu)|$, in which $a_1$ is the normalization parameter, $\nu_0$, $u_g$, and $\tau$ are the LO phonon resonance frequency, group velocity, and lifetime, respectively, in the effective permittivity of monolayer WSe$_2$ (eq. 1), and $T'_{\text{sim/exp}}$ is the simulated or measured relative transmission spectrum of the nanoslot array covered with monolayer WSe$_2$. The fitting parameters to minimize $f'_{\text{obj}}$ were $a_1$, $\nu_0$, $u_g$, and $\tau$. Since the phonon mode feature is located in a relatively narrow region around 7.5 THz, the fitting range was limited to 7.2 – 7.64 THz, to reduce the computational heaviness of the procedure. The following values were retrieved: $a_1 = 2.0851$, $\nu_0 = 7.474$ THz, $u_g = 610$ m/s, and $\tau = 5.56$ ps.

Note that the normalization parameters $a_0$ and $a_1$ allow us to take into account the fact that, in our experiments, the patterned nanoslot array area in the gold film is smaller than the illuminating THz beam. The slight difference between $a_0$ and $a_1$ can be attributed to the non-perfectly identical alignment of the array with respect to the larger THz beam in the two measurements, before and after the transfer of monolayer $WSe_2$. These parameters were used to appropriately compare the experimental relative transmission spectra of the X-shaped nanoslots with and without monolayer $WSe_2$ (grey curves in Figure 6c), which were plotted as $T'_{exp}(\nu)$ and $(a_0/a_1) \cdot T_{exp}(\nu)$, respectively. The simulated spectra shown as red curves in the same figure correspond to $T'_{sim}(\nu)$ and $T_{sim}(\nu)$.

*Fabrication of the X-shaped nanoslot array:* A high-resistivity silicon wafer (10 kΩ·cm) was cut into 10 × 10 mm² samples. Such samples were then cleaned in acetone, isopropyl alcohol (IPA), and water in an ultrasonic bath (1' for each substance) and finally treated in oxygen plasma. The silicon substrate was covered with a 5-nm-thick chrome adhesion layer and a 100-nm-thick gold film via electron beam physical vapor deposition using a Kenosistec KE500ET system. The sample was then spin-coated with an electron beam resist (polymethyl methacrylate-PMMA-A4) at 1800 rpm for 60". Later, the substrate was baked on a hotplate at 180 degrees Celsius for 7'. The lithographic step was performed by using a Raith 150-two electron beam lithography system, and an array area of 2.5 × 2.5 mm² was patterned with the X-shaped nanoslot design. The PMMA layer was then developed in a solution of Methyl isobutyl ketone (MIBK)/IPA (1:3) for 30" at 4 degrees Celsius and stopped in IPA for 20". The patterned PMMA layer was used as a mask for the dry etching process, which was conducted in a SENTECH SI500 plasma-enhanced reactive ion etching system. The process was performed with 40 sccm of Ar gas flow at -5 degrees Celsius, with a reactor pressure of 1 Pa and a RF generator power of 60 W. The process was divided into two consecutive steps of approximately 100" each, to maintain the temperature around -5 degrees Celsius (the exact etching time was determined through scanning electron microscopy inspection of the sample after each step). Finally, the PMMA was removed in hot acetone, a procedure followed by an oxygen plasma treatment.

*Terahertz measurements:* THz transmission spectroscopy was carried out using a Fourier Transform Spectrometer (FTS, Blue Sky Spectroscopy Inc.) equipped with a $Si_3N_4$ blackbody source. A pyroelectric detector (QMC Instruments Ltd.) was employed to collect the transmitted THz radiation during the interferogram scans. The samples with dimensions of 10×10 mm² were mounted on a motorized rotation stage, which was then installed inside the FTS chamber.

To remove water vapor-related absorption lines in the measurements, the FTS chamber was vacuum-pumped for one hour until it reached a pressure value below 100 mTorr. THz transmission spectra were retrieved from an average of 1000 interferograms acquired with a nominal resolution of 0.15 cm$^{-1}$. Before Fourier transforming the interferogram traces, Kaiser-Bessel-derived windows[41] were used to remove from such data echo peaks intrinsic to our spectrometer configuration as well as resulting from Fabry-Pérot reflections associated with the silicon substrate (see Figure S10 in Supporting Information for more details).

*Raman and photoluminescence spectroscopy*: The CVD-grown WSe$_2$ transferred on the nanoslot array was characterized by Raman spectroscopy using a Horiba system (iHR320) equipped with a confocal BX-41 Olympus microscope, a linearly polarized diode-pumped continuous-wave solid-state laser (TEM 00 DPSS laser) (04-01 Cobolt) with a 473 nm wavelength, and a back-illuminated deep depleted charge-coupled device. The double polarized measurements were acquired using a half waveplate (polarizer) and a Glan-Thomson prism (analyzer, extinction ratio 100 000:1). The laser power was set at 5.91 mW and 13.9 mW, respectively, for the polarized and unpolarized Raman measurements (Figure S7). The focal spot size was 0.6 μm, and the acquisition time was 700 s for all Raman measurements. The photoluminescence spectra of monolayer WSe$_2$ were obtained with a laser excitation wavelength of 445 nm and a laser power of 160 mW.

*Density-functional-theory calculations*: The phonon dispersion of monolayer WSe$_2$ was computed using the density functional theory software package Quantum Espresso.[25] The generalized gradient approximation (GGA) of the Perdew-Burke-Ernzerhof (PBE) exchange-correlation functional[42] was employed. The ultrasoft pseudopotentials[43] of tungsten and selenium used in the calculation are available on materialscloud.org. The kinetic energy and density cutoffs were set at 80 Ry and 960 Ry, respectively. The 2D material domain cutoff method was used to truncate the calculation domain.[44] Hubbard correction was implemented using DFPT.[45] Brillouin zone integration was performed using a 12 × 12 × 1 Monkhorst-Pack grid. The atomic structure was relaxed with forces on the atoms smaller than 0.0001 eV/Å. The Raman activity of monolayer WSe$_2$ was calculated using Quantum Espresso with the incorporated local-density approximation (LDA). The pseudopotentials of tungsten and selenium used in the Raman calculation were generated by the code of the Optimized Norm-Conserving Vanderbilt Pseudopotential (ONCVPSP),[46] available on pseudo-dojo.org.

**Acknowledgements**

L.R. and E.O. are grateful for financial support from the Natural Sciences and Engineering Research Council of Canada (NSERC) through the Discovery Grant scheme. A.T. and V.A. would like to acknowledge the support from the European Research Council (ERC) under the European Union's Horizon 2020 research and innovation program "REPLY ERC-2020-COG Grant agreement No. 101002422. X.J. would like to thank the Fonds de Recherche du Québec − Nature et Technologies (FRQNT) for a postdoctoral fellowship (B3X-288650). X.J., Y-G.J., L.R., and P.B. acknowledge CMC Microsystems for software licensing and Digital Research Alliance of Canada – Resources for Research Groups (RRG) 2021(ID3581) & 2023 (ID4456) for the large-resource computational services. The authors would like to thank Artiom Skripka and Fiorenzo Vetrone for their help with the photoluminescence measurements. X.J. would also like to thank Zimin Feng for DFT software training and related discussions.